# Understanding Molecular Basis of PTPN11-Related Diseases

**Seungha Um**

National Human Genome Research Institute, National Institutes of Health, Bethesda, MD, USA,
seungha.um@nih.gov

**Tulika Kakati**

Department of Psychiatry, University of
California San Diego, La Jolla, CA, USA,
tkakati@health.ucsd.edu

**Lilia M Iakoucheva**

Department of Psychiatry, University of
California San Diego, La Jolla, CA, USA
lilyak@ucsd.edu

**Yile Chen**

Department of Biomedical Informatics and
Medical Education, University of
Washington, Seattle, WA, USA,
yilechen@uw.edu

**Sean Mooney**

National Human Genome Research Institute, National Institutes of Health, Bethesda, MD, USA,
sean.mooney@nih.gov

**Abstract**

The PTPN11 gene encodes the Src homology 2 domain-containing protein tyrosine phosphatase (SHP2), a key regulator of cell growth, differentiation, and apoptosis through its modulation of various signaling pathways, including the RAS/MAPK signaling pathway. Missense variants in PTPN11 disrupt SHP2's proper catalytic activity and the regulation of signaling pathways, leading to disorders such as Noonan syndrome (NS), LEOPARD syndrome (LS), or juvenile myelomonocytic leukemia (JMML). These missense variants have molecular disruptions resulting in gains and losses of function at both the molecular and phenotypic levels. Depending on their location within SHP2, missense substitutions disrupt inter-domain regulation or impair phosphatase function, resulting in altered phosphatase activity. In this study, we investigate the molecular basis underlying the differential pathogenicity of PTPN11 missense variants and predict the structural consequences of these variants using MutPred2 and AlphaFold2. We find that LOF and GOF variants display distinct functional mechanisms in sodium and DNA binding, and that NS-associated missense variants identified in fetuses with ultrasound-detected anomalies and familiar cases are more likely to be pathogenic.

*Keyword: PTPN11, SHP2, pathogenicity, MutPred2, AlphaFold2*

# 1. Introduction

The gene PTPN11 (OMIM 176876), located on chromosome 12q24, encodes SHP2 protein. The SHP2 protein is non-receptor protein-tyrosine phosphatase (PTP) composed of 593 amino acids. SHP2 comprise two tandemly arranged amino-terminal Src-homology 2 (SH2) domains (N-SH2 and C-SH2), a PTP catalytic domain, and a C-terminal tail containing two main tyrosine phosphorylation sites (Tyr542 and Tyr580) (Neel et al., 2003). Both the N-SH2 and C-SH2 domain of SHP2 selectively bind to short amino acid motifs containing a phosphotyrosyl residue, which regulate the catalytic activity of SHP2. SHP2 is critical for signal transduction through the RAS/MAPK signaling pathway, linking growth factor signaling to RAS activation and thereby regulating a wide range of cellular processes, including cell growth, proliferation, and differentiation. SHP2's catalytic activity acts as a positive regulator of the RAS/MAPK signaling pathway, where its phosphatase activity promotes pathway activation (Neel et al., 2003; Barford & Neel, 1998; Hof et al., 1998). This activity is tightly controlled through its recruitment by binding partners. Also, it has been found that SHP2 modulates the phosphoinositide-3 kinase (PI3K)/AKT cascades (Tajan et al., 2015).

Missense variants in SHP2 result in aberrant regulation of the RAS/MAPK signaling pathway, leading to profound effects associated with several human diseases, including developmental disorders such as LEOPARD syndrome (LS) and Noonan syndrome (NS), as well as cancers such as juvenile myelomonocytic leukemia (JMML). NS exhibits autosomal dominant inheritance characterized by facial dysmorphism including hypertelorism, low-set ears and ptosis, short stature, skeletal abnormalities, and heart defects (Allanson, 1987; Mendez et al., 1985). NS is a relatively common Mendelian disorder with a panethnic prevalence of 1:1,000 to 1:2,500 live births (Roberts et al., 2013; Allanson, 1987). Germline missense mutations in PTPN11 were found in 45% of clinically diagnosed Noonan syndrome patients (Tartaglia et al., 2001). NS significantly contributes to congenital heart disease, typically presenting with pulmonary valve dysplasia and hypertrophic cardiomyopathy (Roberts et al., 2013; Kontaridis et al., 2006)

LS is also an autosomal dominant developmental disorder and the term "LEOPARD" reflects the characteristic clinical features of the disorder: multiple lentigines (L), electrocardiographic conduction abnormalities (E), ocular hypertelorism (O), pulmonary stenosis (P), abnormalities of genitalia (A), retardation of growth (R), and deafness (D). LS and NS share several overlapping features since both are RASopathies which are a group of autosomal dominant disorders caused by variants in genes that affect the RAS/MAPK signaling pathway. These RASopathies present phenotypical overlap with common clinical features resembling NS. However, multiple lentigines are rarely observed in NS (Roberts et al., 2013) and this difference has led to LS sometimes being referred to as "Noonan syndrome with multiple lentigines".

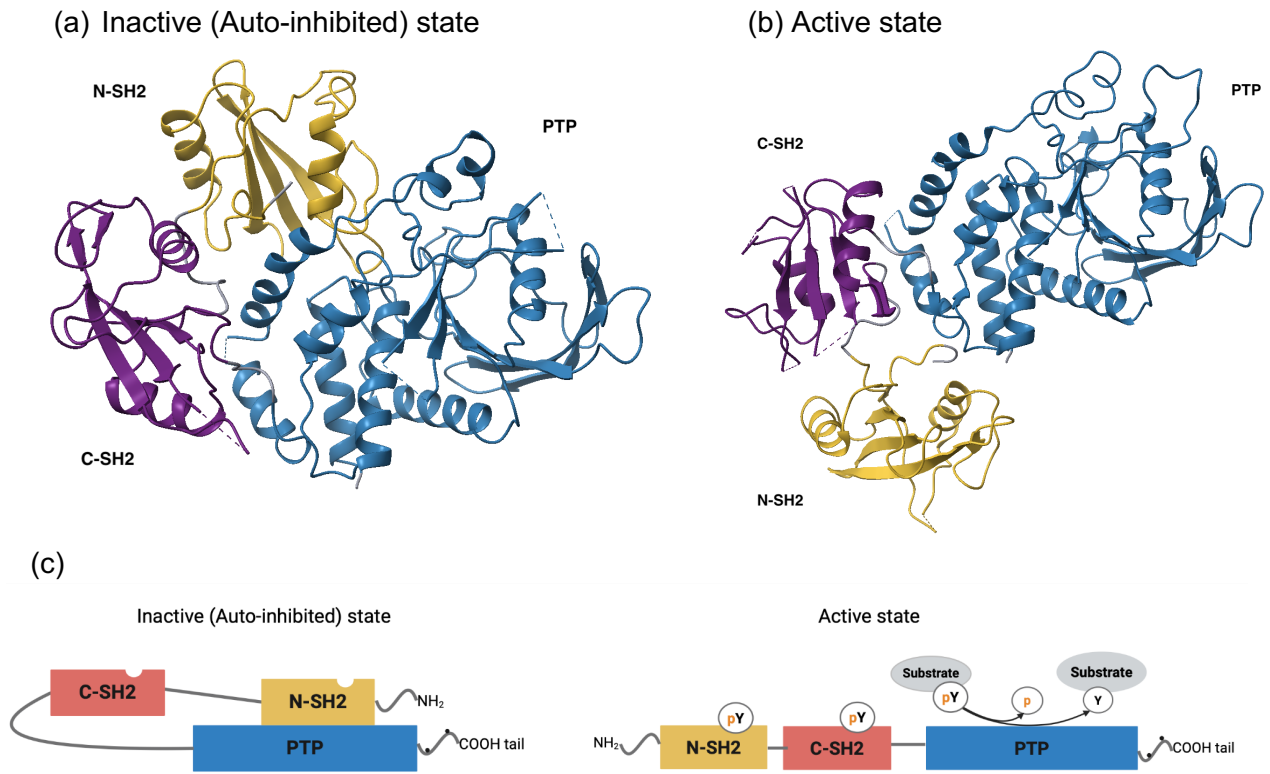

*Figure 1. The overall structure of SHP-2. The N-SH2, C-SH2 and protein tyrosine phosphatase (PTP) domains are shown in yellow, purple and blue, respectively. (a) The auto-inhibited (closed) state of wild type of SHP2 (PDB ID: 2SHP), (b) The active (open) state of SHP2 with E76K variant (PDB ID: 6CRF), (c) Schematic diagram of SHP2 structure for each state. Dots in the C-terminal tail represent phosphorylation sites (Tyr542 and Tyr580)*

JMML is a myeloproliferative disorder characterized by excessive proliferation of myelomonocytic cells, and somatic PTPN11 variants have been identified in 34% of JMML cases as well as in a small percentage of myelodysplastic syndrome, acute myeloid leukemia, and acute lymphoblastic leukemia cases in patients without NS (Tartaglia et al., 2003; Loh, 2004; Tartaglia, 2004). JMML arises sporadically in individuals without NS who harbor somatic PTPN11 missense variants and, less commonly, in patients with NS carrying germline PTPN11 variants (Tartaglia et al., 2003). This association between NS and JMML reflects the difference in oncogenic strength: NS-associated PTPN11 variants act as relatively weak drivers of proliferation, a hallmark of cancer, whereas the somatic variants seen in JMML have stronger molecular defects and would likely be embryonically lethal if present in the germline (Liu et al., 2023). This strength difference was examined through phosphatase activity association analyses, implying that JMML-associated variants impair SHP2 phosphatase activity more severely than NS-associated variants (Tartaglia et al., 2003; Niihori et al., 2005).

The phenotypic variability of PTPN11 variants is attributed to the locations of the substitutions, resulting in differential phosphatase activation. To investigate the distinct molecular mechanisms underlying the diverse phenotypes associated with missense variants in PTPN11, we perform an in silico analysis using MutPred2, a machine learning model that predicts the functional and structural impact of amino acid substitutions. By applying MutPred2, we assess how PTPN11 missense variants alter protein structure and function in ways that contribute to distinct clinical outcomes. We also investigated whether additional clinical features or patterns of inheritance observed in individuals with NS are associated with an increased likelihood of pathogenicity. To further explore the functional consequences of PTPN11 missense variants, we assessed their structural impact using AlphaFold2 and examined the tissue-specific expression patterns of PTPN11 isoforms. This integrative approach highlights the utility of variant effect prediction tools in revealing potential pathogenic mechanisms, prioritizing candidate variants, and generating hypotheses for future experimental validation.

## 2. Structure and signaling of SHP2

SHP2 is a ubiquitously expressed cytoplasmic PTP encoded by PTPN11. Its N-terminal and C-terminal SH2 domains selectively bind short phosphotyrosine-containing motifs, which regulate SHP2's catalytic activity. In addition, tyrosine phosphorylation sites within its C-terminal tail have been shown to serve as docking sites for adaptor proteins with SH2 domains, such as GRB2, thereby linking SHP2 to downstream signaling pathways (Tartaglia et al., 2001; Kontaridis et al., 2006). In the inactive (auto-inhibited) state, the N-SH2 domain, specifically the side opposite to its phosphotyrosyl peptide binding pocket, interacts with the PTP domain (Figure 1 (a)). This intramolecular interaction of SHP2 results in autoinhibition of its PTP activity preventing substrate access to the active site (Hof et al., 1998). While C-SH2 domain is not directly involved in the auto-inhibitory interaction like N-SH2 domain, C-SH2 domain recognizes and binds phosphotyrosine motifs on interacting proteins. This interaction not only guides the proper localization of SHP2 within cellular signaling complexes but also supports stabilization of its active conformation upon relief of autoinhibition. Upon binding of its tandem SH2 domains to phosphotyrosyl protein ligands (e.g., scaffolding adaptors), the autoinhibitory interface is disrupted, releasing the PTP domain and triggering phosphatase activation (Figure 1(b)). This conformational change exposes the catalytic site, thereby activating the phosphatase to act on its substrates. SHP2 plays a crucial role in maintaining the cellular equilibrium of protein tyrosine phosphorylation through its coordinated action with protein tyrosine kinases (PTKs). Disruption of the balance between protein tyrosine kinases and phosphatases, leading to abnormal tyrosine phosphorylation, has been implicated in various human diseases, including cancer, developmental disorders, and autoimmune disorders (Kontaridis et al., 2006; Tartaglia et al., 2002).

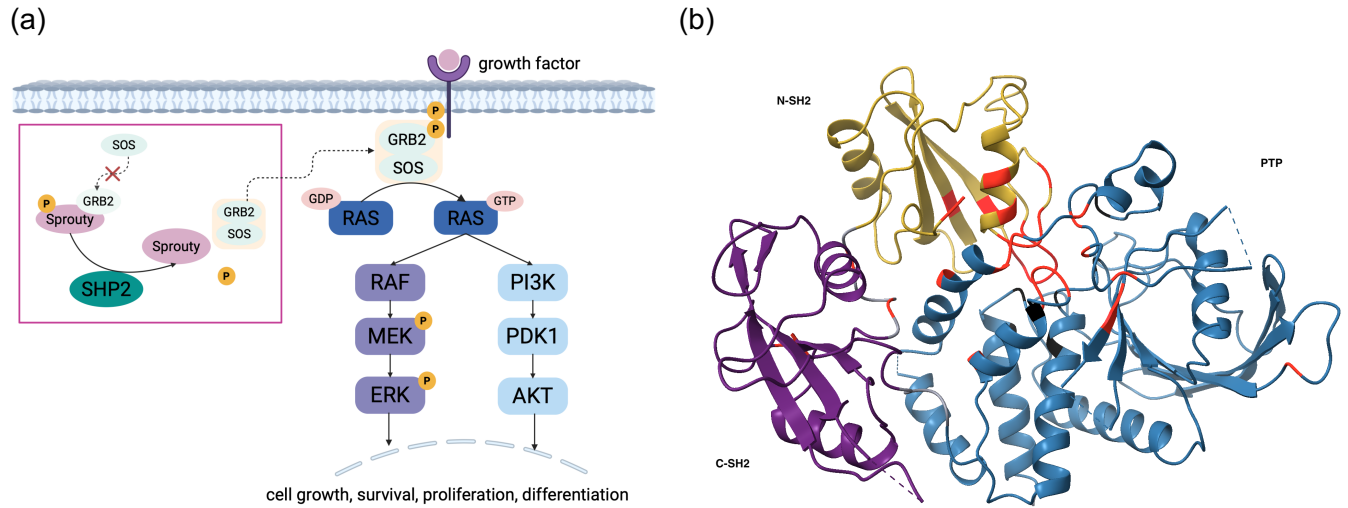

*Figure 2. (a) SHP2-associated growth factor dependent signaling pathways, including RAS/MAPK and PI3K/AKT pathway. (b) SHP2 crystal structure (PDB: 2SHP) highlighting NS/JMML-associated variants in red and LS-associated variants in black. The domains are colored N-SH2 (yellow), C-SH2 (purple), and PTP (blue)*

Based on conformational change which controls the activation, SHP2 positively regulates the RAS/MAPK pathway downstream of RTKs and cytokine receptors by facilitating RAS activation (Figure 2(a)). The RAS/MAPK pathway is a critical signal transduction cascade that mediates cellular responses—such as proliferation, differentiation, survival, and metabolism—in response to extracellular signals, including growth factors, cytokines, and hormones (Chang & Karin, 2001). Ligand binding to cell surface receptors induces phosphorylation within their cytoplasmic domains, promoting the recruitment of adaptor proteins like GRB2. These adaptors form complexes with guanine nucleotide exchange factors such as SOS, facilitating the activation of RAS by promoting the exchange of GDP for GTP. Active RAS then initiates a series of phosphorylation events involving RAF, MEK, and ERK. Once activated, ERK moves into the nucleus to influence gene expression and also acts on targets in the cytoplasm, coordinating both short-term and long-term cellular responses (Roberts et al., 2013). SHP2 facilitates this process by dephosphorylating inhibitory sites and stabilizing protein-protein interactions within the signaling complex, thereby promoting efficient signal propagation. SHP2's catalytic activity acts as a positive regulator of the RAS/MAPK pathway, where its phosphatase activity promotes pathway activation. Also, it has been found that SHP2 modulates the PI3K/AKT cascades (Tajan et al., 2015). Missense variants in SHP2 result in aberrant regulation of the RAS/MAPK pathway driving uncontrolled cell proliferation and differentiation and developmental disorders such as NS as well as various leukemias.

## 3. Missense variants in PTPN11 and associated functional impact

Missense variants in the PTPN11 gene disrupt normal regulation of the RAS/MAPK pathway and these variants are associated with several human diseases, including LS (Maheshwari et al., 2002; Kontaridis et al., 2006), NS (Tartaglia et al., 2001; Maheshwari et al., 2002) and JMML (Tartaglia et al., 2003). The clinical phenotypes are largely determined by the missense variant's location, which affects how SHP2 activity is dysregulated. Specifically, variants impacting phosphatase activity are primarily located in two regions: (i) at the interacting surfaces between the N-SH2 and PTP domains, resulting in increased phosphatase activity, and (ii) within the PTP domain but not at the interacting surfaces, leading to decreased phosphatase activity. Missense variants associated with LS, NS, and JMML are mapped onto the SHP2 structure shown in Figure 2(b).

Most PTPN11 missense variants associated with NS or JMML are found at the interacting surfaces between the N-SH2 and PTP domain, where they disrupt SHP-2's autoinhibitory conformation. This alteration shifts SHP-2 into a more active state, enhancing its phosphatase activity and leading to excessive activation of the RAS/MAPK pathway, thereby producing gain-of-function (GOF) effects. Disruptions in SHP2's conformational switching mechanism are central to the development of NS and JMML. Although both NS and JMML are associated with GOF effects, JMML associated variants exhibit significantly more pronounced molecular defects than those observed in NS (Tartaglia et al., 2003). Somatic variants identified in JMML display stronger molecular disruptions that could cause embryonic lethality if present in the germline. In contrast, the molecular defects associated with NS variants are milder and insufficient to drive malignant transformation. Based on these observations, previous studies examine the association between phosphatase activity levels and embryonic lethality (Niihori et al., 2005).

LS-associated variants have been identified solely in the PTP domain, impairing its phosphatase activity producing loss-of-function (LOF) effects. This impaired phosphatase activity by missense variant leads to reduced phosphatase activity even in the open state. The enzymatic properties of these variants show a dominant-negative effect, where the inactive variant SHP2 interferes with growth factor/ERK-MAPK mediated signaling, despite the presence of one functional copy of the gene (Kontaridis et al., 2006). While LOF-associated variants are usually recessive because one functional copy of the gene often suffices, LS-associated variants show a dominant-negative effect. In these cases, the inactive SHP2 variant interferes with growth factor–induced RAS/MAPK signaling, despite the presence of one functional copy of the gene (Kontaridis et al., 2006).

## 4. Methods

### 4.1 Pathogenicity and molecular basis prediction

MutPred2 is a machine learning model that predicts the impact of variants on specific aspects of protein structure and function (Pejaver et al., 2020). This algorithm quantifies the pathogenicity of amino acid substitutions and predicts their impact on phenotype by modeling a wide range of structural and functional changes based on amino acid sequences. MutPred2 models changes across more than 50 structural and functional protein properties, including secondary structure, signal peptide and transmembrane topology, catalytic activity, macromolecular binding, PTMs, metal binding and allostery. It also provides a posterior probability that quantifies the loss or gain of a property due to an amino acid substitution, offering a clear interpretation of the molecular consequences. MutPred2 provides an interpretable pathogenicity score ranging from 0 to 1, with higher scores indicating a greater likelihood of pathogenicity. Applying a machine learning model is invaluable because it not only identifies variants that may alter phenotypes but also elucidates the relationship between altered molecular mechanisms and clinical features. Consequently, it guides further research, especially when experimental techniques such as multiplexed assays of variant effect (MAVEs) are cost-prohibitive and labor-intensive.

### 4.2 Assessment protein structure changes from substitutions

AlphaFold2 (https://alphafoldserver.com/) is used to predict the structural effects of PTPN11 missense variants. We compared the predicted variant structures with the wild-type SHP2 structure. To quantify how each substitution alters distances between residue pairs, we construct distance matrices for both wild-type($D^{WT}$) and variant ($D^{variant}$) structures. Each element of the distance matrix is defined as

$$d_{ij} = ||r_i - r_j||$$

where $r_i$ and $r_j$ denote the 3D Cartesian coordinates of the the 3D Cartesian coordinates of the $C_\alpha$ atoms of residues $i$ and $j$ ($i, j = 1, \ldots, 524$). Although the full SHP2 chain spans 593 residues, we exclude the C-terminal tail (residues 525–593) due to high prediction uncertainty, thereby focusing on the structured domains. We then compute the element-wise absolute difference matrix

$$\Delta = |D^{WT} - D^{variant}|$$

where each element quantifies the change in $C_\alpha - C_\alpha$ distance between residues $i$ and $j$ induced by the substitution.

## 5. Result

### 5.1 Prediction of pathogenicity and molecular basis of missense variants in PTPN11

By applying MutPred2, we investigate the alterations in local structural and functional properties induced by GOF and LOF variants, and to examine how JMML associated variants, which exhibit more pronounced molecular defects than those found in NS—affect these properties. We further seek to explore the relationship between variants that manifest with severe phenotypes accompanied by additional clinical features and their corresponding pathogenicity scores.

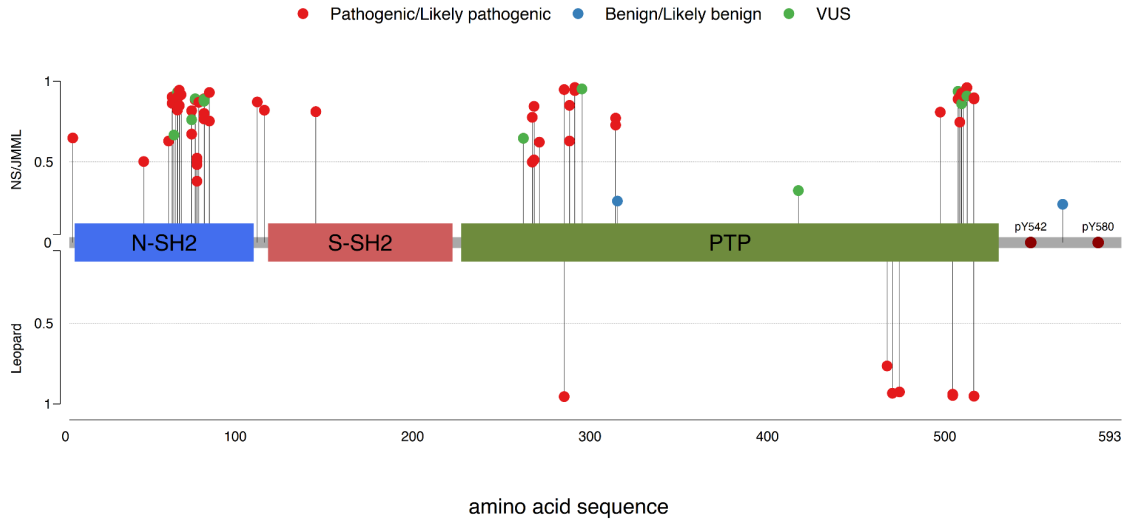

*Figure 3. Schematic representation of the SHP-2 protein, illustrating its functional domains along with tyrosine phosphorylation sites (Y542 and Y580) in the C-terminal hydrophobic tail. The line length is proportional to the MutPred2 pathogenicity score for the corresponding variant (ranging from 0 to 1), with higher scores indicating a greater likelihood of pathogenicity. Each dot indicates the location of a missense variant, with colors corresponding to ClinVar pathogenicity classifications: red for pathogenic/likely pathogenic, blue for benign/likely benign, and green for conflicting or uncertain significance. The upper and lower panels illustrate the pathogenicity scores for variants associated with NS/JMML and LS, respectively.*

First, we evaluate MutPred2 pathogenicity scores for all missense variants in PTPN11 reported to be associated with NS, JMML, or LS that have received at least one review star in ClinVar. The list of these variants is provided in Supplementary Table 1 along with predicted pathogenicity scores. In Figure 3, schematic representation of the SHP-2 protein is represented and different dot colors represent ClinVar pathogenicity classifications: red for pathogenic/likely pathogenic, blue for benign/likely benign, and green for conflicting or uncertain significance. The line length is proportional to the MutPred2 pathogenicity score, with higher scores indicating a greater likelihood of pathogenicity. The upper and lower panels illustrate the pathogenicity scores for variants associated with NS/JMML and LS, respectively. Since these variants are associated with disorders, most are classified as pathogenic in ClinVar and predominantly exhibit MutPred2 pathogenicity scores above 0.5. Two variants, I309V and L560F, are classified as benign/likely benign despite their reported association with NS/JMML. This classification is primarily due to their relatively high allele frequencies, which exceed the thresholds defined by ClinVar for PTPN11 pathogenicity. Notably, both variants also have MutPred2 scores below 0.5, confirming that MutPred2 scores align better with the theoretical expectations of allele frequencies for slightly deleterious variants. Although three pathogenic variants (A72S, A72T, and L261F) have pathogenicity scores below 0.5 and warrant further investigation, MutPred2 provides effective predictions for the pathogenicity of PTPN11 variants.

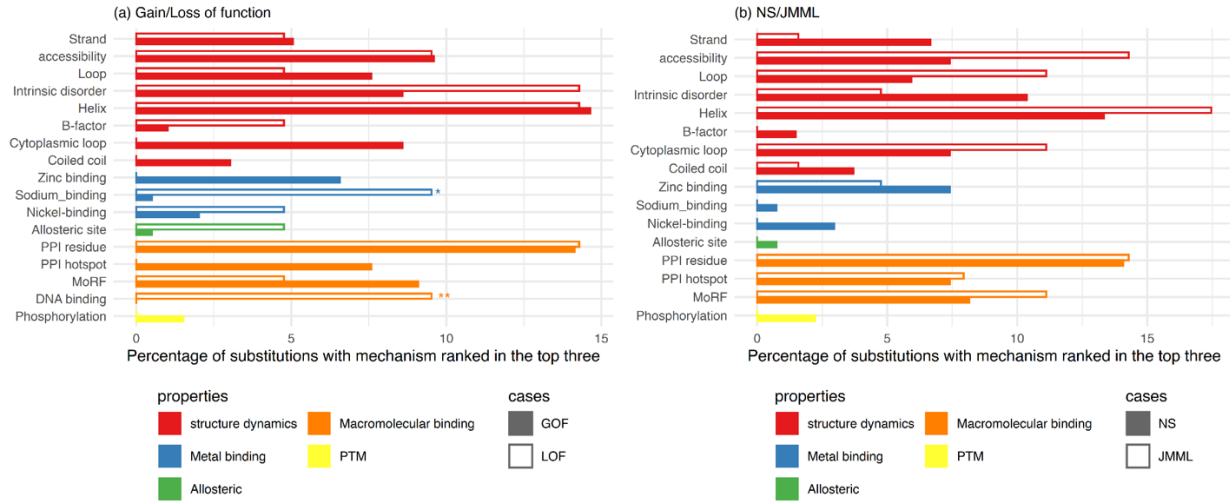

*Figure 4. Enrichment of structural and functional signatures of GOF variants versus LOF variants (a) and NS versus JMML (b). Properties are grouped based upon their broader classes as described in the MutPred2 ontology. Statistical significance was assessed using a two-sided Fisher's exact test. Asterisks indicate significance: $0.01 < p < 0.05$ (*), $p < 0.01$ (**).*

Next, we investigate whether variants associated with PTPN11 disorders preferentially impact specific protein structural and functional properties, given that alterations in protein structure are a common mechanism underlying Mendelian disorders. In particular, we identify which molecular mechanisms are most frequently ranked among the top three for both GOF and LOF variants. Based on these top-ranked mechanisms for each variant, we plot the frequency distribution in Figure 4(a). Both GOF and LOF variants affect high proportions of structure dynamics (strand, accessibility, Loop, intrinsic disorder, helix, and B-factor, cytoplasmic loop and coiled coil) suggesting that alterations in protein structure are common mechanisms underlying missense variants in PTPN11. On the other hand, LOF variants show significant enrichment for sodium binding (one of metal binding) and DNA binding (macromolecular binding) compared to GOF variants, with statistically significant differences. We also examine the frequency distribution based on these top-ranked mechanisms for NS- and JMML-associated variants in Figure 4(b). We observed an enrichment of certain properties—such as Molecular Recognition Features (MoRFs), solvent accessibility, and helical propensity—among JMML associated variants. However, these differences were not statistically significant when compared to NS-associated variants. Although JMML-associated variants are generally thought to have stronger molecular defects, our findings do not provide support for this distinction. This may be due in part to the limited number of currently known JMML associated variants and the presence of overlapping variants in both NS and JMML cases. Additionally, the strength of molecular defects is not directly captured by this frequency distribution. If more distinct variants are identified in the future, they may help clarify whether the stronger molecular defects reflect true mechanistic differences.

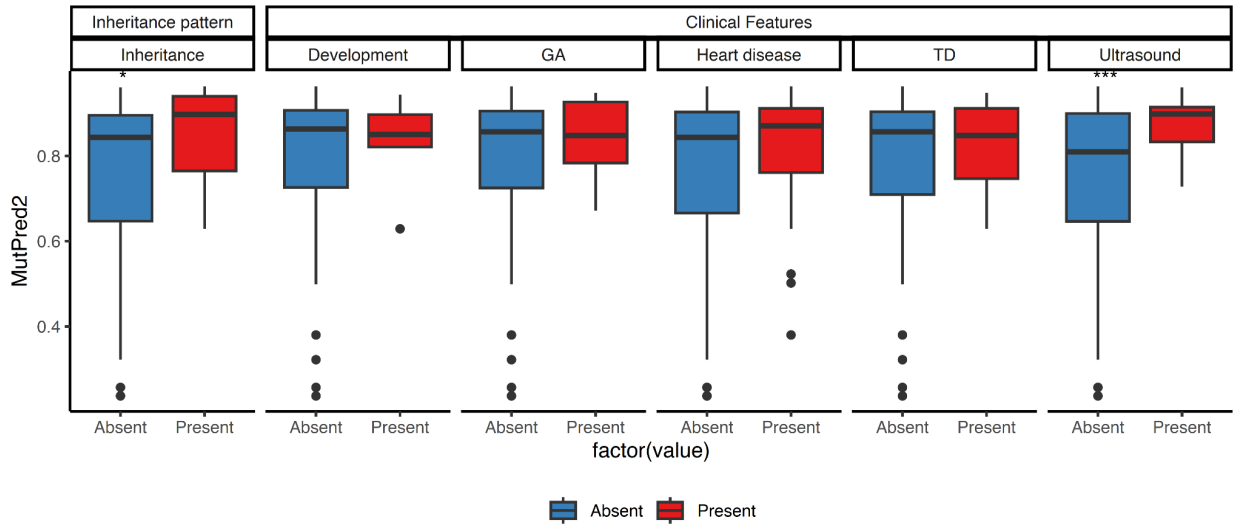

*Figure 5. Boxplot of pathogenicity scores stratified by clinical features and inheritance patterns. Colors indicate the presence or absence of each feature or pattern. TD = Thorax deformity, GA = Genitourinary anomalies, Ultrasound = fetal anomalies detected via ultrasound in cases of NS, such as cystic hygroma and increased nuchal translucency (NT). Asterisks ∗∗ indicate statistical significance (p-value ≤ 0.01).*

Along with the spectrum of PTPN11 variants, the phenotypic diversity corresponding to NS-associated variants has been extensively investigated. NS patients exhibit varying degrees of additional clinical features, including heart disease(Atik et al., 2016; Ezquieta et al., 2012), thoracic deformities (Atik et al., 2016), developmental delays (Atik et al., 2016) and abnormalities detectable via fetal ultrasound, such as cystic hygroma and increased nuchal translucency (Leach et al., 2019). Furthermore, studies have examined whether the identified variants are de novo or familial cases (Atik et al., 2016; Zepeda-Olmos et al., 2024). In this study, we examine whether additional clinical features or patterns of inheritance observed in NS patients are associated with pathogenicity scores. Our premise is that variants accompanied by clinical features commonly observed in individuals with NS, or found in familial cases, are more likely to be pathogenic. As shown in Figure 5, NS variants associated with ultrasound-detectable anomalies and familial cases exhibit a higher likelihood of pathogenicity. The differences in both ultrasound findings and familial cases are statistically significant, suggesting that variants identified through early diagnostic indicators or family history tend to be more pathogenic.

## 5.2 Protein structure prediction with AlphaFold2

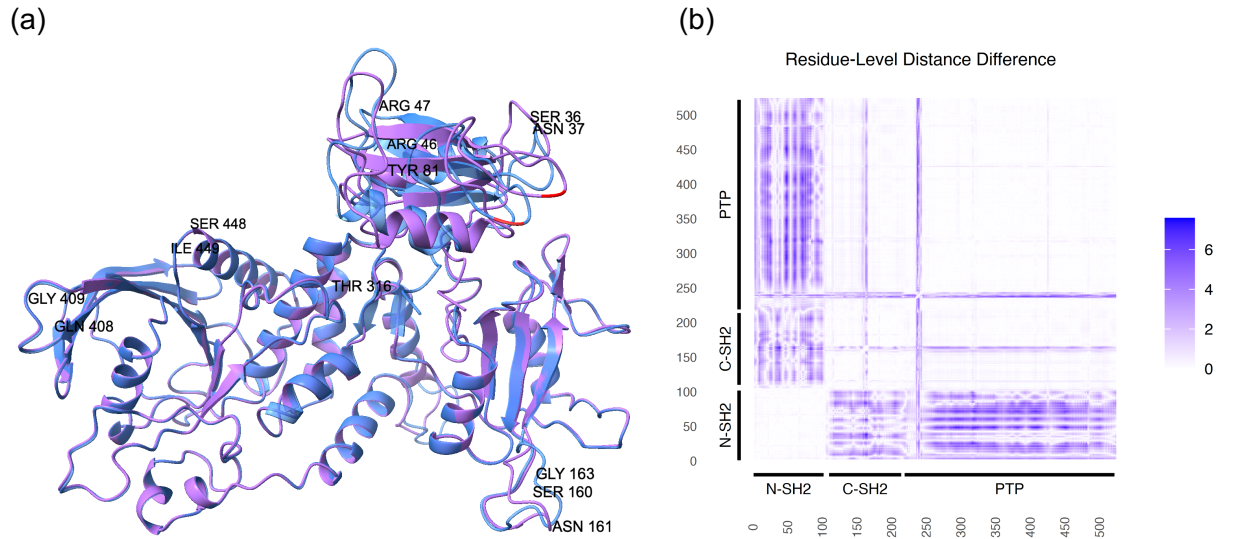

*Figure 6. (a) SHP2 predicted structures with AlphaFold2 with variant D61N (blue) and wild type PTPN11 (pink). 61 residue is indicated by red. (b) heatmap of absolute difference in distances between residue pair C-C distance between wild type SHP2 and D61N.*

We compare the wild-type SHP2 structure with the D61N (NS-associated) variant structure predicted by AlphaFold2 (Figure 6(a)). To quantify structural changes, we computed residue–residue distance matrices for the wild type ($D^{WT}$) and D61N ($D^{D61N}$), and then calculated the element-wise absolute difference:

$$\Delta = |D^{WT} - D^{D61N}|$$

where each element quantifies the change in $C_\alpha - C_\alpha$ distance between residues $i$ and $j$ induced by the D61N substitution. A heatmap visualizing the pairwise distance changes $\Delta$, is shown in Figure 6(b). The top five largest distance differences are annotated on the structure in Figure 6(a) and listed in Supplementary Table 2. As shown in the figure and table, D61N induces local perturbations in the N-SH2 domain, altering distances between N-SH2 residues and those in other domains. It is important to note that AlphaFold2 predicts protein structures in a static, likely inactive conformation. Therefore, the impact of the variant on the self-inhibitory conformation and the resulting constitutive activation of the phosphatase cannot be assessed using AlphaFold2, as such effects are governed by protein dynamics not captured in static structural predictions. Nevertheless, the observed changes in pairwise residue distances may reflect altered interdomain interactions that could disrupt the self-inhibitory conformation and ultimately modulate phosphatase activity.

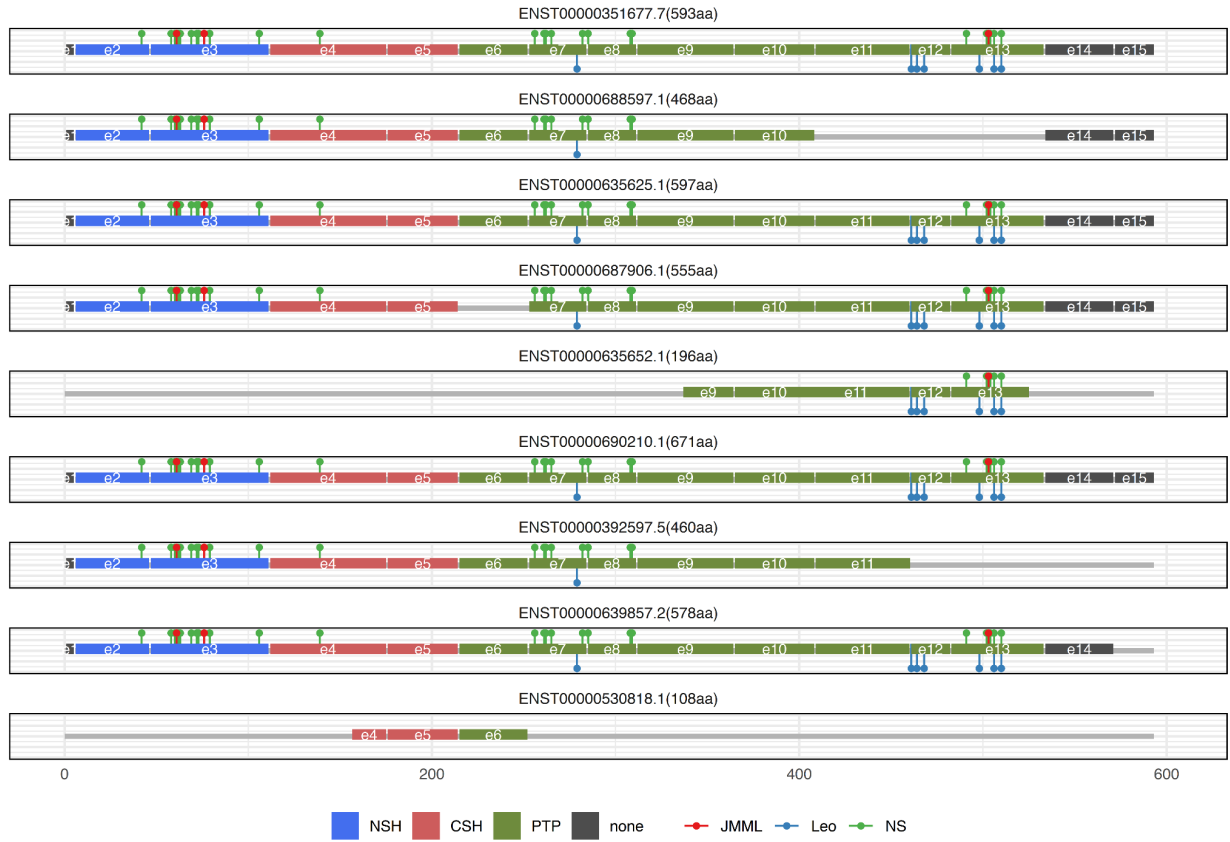

*Figure 7. Exon structures of the nine PTPN11 isoforms with open reading frames (ORFs). Disease-associated variants are annotated in distinct colors, and protein domains are highlighted in different colors.*

## 5.3 Tissue Expression of PTPN11 Isoforms

The PTPN11 gene, located on chromosome 12, comprises 16 exons and encodes multiple transcript isoforms. The conventional human isoform (ENST00000351677.7) harbors pathogenic variants linked to NS (exons 2, 3, 4, 7, 8, 13), LS (exons 7, 12, 13), and JMML (exons 3, 13). Nine isoforms with open reading frames (ORFs) are selected and visualized with their exon structures, and disease-associated variants are annotated using distinct colors (Fig 7). Among the shorter isoforms, ENST00000688597.1 and ENST00000392597.5 predominantly carry GOF (NS and JMML) variants, while ENST00000635652.1 encodes only the PTP domain. To examine isoform-specific expression across human tissues, we utilize data from the Genotype-Tissue Expression (GTEx) project, a comprehensive resource for investigating tissue- and cell-specific gene expression and regulation. This analysis provides insight into the tissue-specific expression patterns of PTPN11 isoforms and their potential relevance to disease mechanisms.

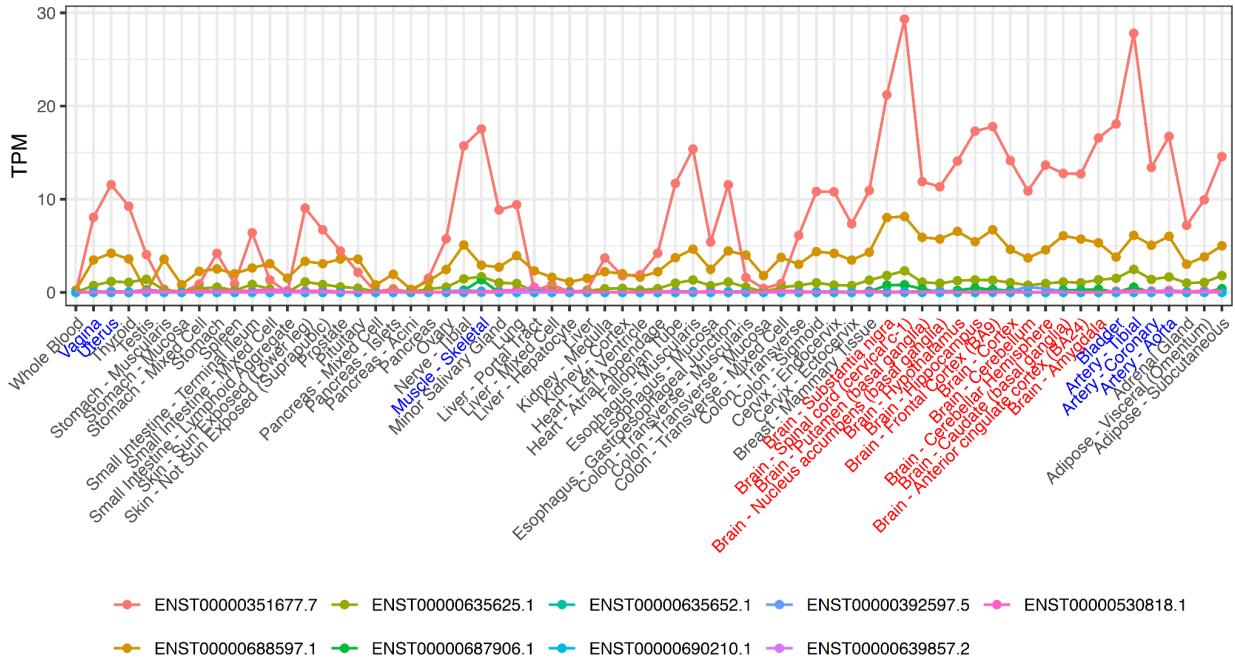

*Figure 8. Transcript expression levels (Transcripts Per Million, TPM) across various human tissues are shown for the nine PTPN11 isoforms. In the x-axis labels, brain tissues are highlighted in red, and tissues associated with clinical features observed in Noonan syndrome (NS) are highlighted in blue.*

In Figure 8, transcript expression levels (Transcripts Per Million, TPM) across various human tissues are shown for the nine PTPN11 isoforms. Three isoforms (ENST00000351677.7, ENST00000688597.1, and ENST00000635625.1) are dominantly expressed across multiple tissues, particularly in brain-associated tissues. The high TPM observed in brain tissues is notable, as developmental delays, learning disabilities, and intellectual disability are frequently reported in individuals with NS (Atik et al., 2016). This suggests that pathogenic PTPN11 variants may disrupt normal expression in brain cells, potentially contributing to neurodevelopmental impairment.

Furthermore, high expression of PTPN11 isoforms is also observed in tissues associated with additional clinical features described in Section 4.1. For instance, genitourinary anomalies may occur with abnormalities in uterine and vaginal tissues; thoracic deformities may involve skeletal muscle abnormalities; and the prenatal ultrasound markers may indicate underlying arterial tissue defects. All of these tissues show elevated TPM levels compared to others, supporting the hypothesis that PTPN11 variants preferentially affect tissues with higher expression, thereby increasing the likelihood of secondary clinical features in individuals with NS.

A comparison between ENST00000688597.1, which contains the majority of NS-associated variants, and ENST00000635652.1, enriched for LS-associated variants, may help elucidate differences in clinical presentation by examining their distinct tissue-specific expression patterns in GTEx data. However, ENST00000635652.1 exhibits minimal expression across most tissues, and NS and LS share overlapping phenotypes due to their common involvement in the RAS/MAPK signaling pathway. Therefore, distinguishing between the two conditions based solely on isoform-specific expression may be limited.

## 6. Conclusions

Missense variants in SHP2 disrupt its catalytic activity and regulation of intracellular signaling pathways, contributing to disease development. In this study, we investigated how these variants cause molecular disruptions leading to GOF and LOF effects at both the pathway and phenotypic levels. These functional outcomes are influenced by the variant's location; variants associated with NS and JMML often occur at the interface between the N-SH2 and PTP domains, disrupting SHP2's autoinhibitory conformation and resulting in hyperactivation of the RAS/MAPK pathway. In contrast, LS-associated variants are found exclusively within the PTP domain, where they impair phosphatase activity and produce LOF effects.

To further explore the molecular basis of these differences, we used MutPred2 to compare the predicted impact of GOF and LOF variants on SHP2 structure and function. Interestingly, LOF variants were more likely to disrupt sodium binding and DNA interaction mechanisms, suggesting distinct molecular consequences compared to GOF variants. We also assessed whether variants accompanied by clinical features commonly observed in NS patients are more likely to be pathogenic. Our analysis showed that NS-associated variants identified through familial cases or early diagnostic indicators—such as fetal ultrasound anomalies—tend to exhibit a higher likelihood of pathogenicity. To examine how PTPN11 missense variants affect SHP2 structure, we used AlphaFold2 to predict local structural changes. Variants that disrupt SHP2's autoinhibitory conformation were shown to induce localized perturbations in the N-SH2 domain, particularly altering residue distances between N-SH2 and other domains, consistent with a destabilization of the closed conformation. We also investigated the expression levels of nine PTPN11 isoforms across multiple tissues, suggesting that clinical features commonly observed in individuals with NS are associated with tissues exhibiting high expression levels.

Given that experimental characterization of all possible variants is not feasible due to time and resource constraints, our in silico analyses provide valuable insights for prioritizing disease-relevant variants. By modeling the effects of specific variants on local protein structure, function, and pathogenicity, we demonstrate how predictive tools can inform future research and guide experimental validation. Our study highlights how variant effect prediction tools can be used to disentangle the molecular mechanisms underlying variant function and offers a framework for prioritizing uncharacterized variants based on their structural and functional impacts. This approach contributes to a deeper understanding of the molecular basis of genetic disorders and supports the development of hypotheses for downstream functional studies.

## Acknowledge

An earlier version of this work was presented as a poster at the NIH Artificial Intelligence Symposium, May 16th, 2025.

## Data Availability Statement

The data supporting this study can be accessed through the ClinVar database (https://www.ncbi.nlm.nih.gov/clinvar/), the Genome Aggregation Database (gnomAD) (https://gnomad.broadinstitute.org/), and the Genotype–Tissue Expression (GTEx) Project (https://gtexportal.org/).

## Funding statement

This research was supported by the Intramural Research Program of the National Institutes of Health (NIH). The contributions of the NIH author(s) are considered Works of the United States Government. The findings and conclusions presented in this paper are those of the author(s) and do not necessarily reflect the views of the NIH or the U.S. Department of Health and Human Services.

## Conflict of interest

The authors declare no conflicts of interest.

## Supplementary materials

| variant | disease | MutPred2 score |
|---|---|---|
| T2I | NS | 0.648 |
| T42A | NS | 0.502 |
| I56V | NS | 0.629 |
| N58D | NS | 0.863 |
| N58K | NS | 0.902 |
| T59A | NS | 0.666 |
| G60A | NS | 0.875 |
| D61Y | NS | 0.932 |
| D61G | NS | 0.903 |
| D61N | NS | 0.821 |
| Y62C | NS | 0.848 |
| Y62D | NS | 0.944 |
| Y63C | NS | 0.916 |
| E69Q | NS | 0.672 |
| F71I | NS | 0.891 |
| F71L | NS | 0.884 |
| A72V | NS | 0.505 |
| A72S | NS | 0.380 |
| A72G | NS | 0.523 |
| T73I | NS | 0.869 |
| E76A | NS | 0.876 |
| E76D | NS | 0.766 |
| Q79R | NS | 0.753 |
| Q79P | NS | 0.930 |

| Mutation | | |
|---|---|---|
| D106A | NS | 0.871 |
| E110A | NS | 0.820 |
| E139D | NS | 0.811 |
| Q256R | NS | 0.646 |
| L261F | NS | 0.499 |
| L261H | NS | 0.776 |
| L262F | NS | 0.512 |
| L262R | NS | 0.844 |
| R265Q | NS | 0.622 |
| Y279C | NS | 0.948 |
| I282M | NS | 0.850 |
| I282V | NS | 0.629 |
| F285S | NS | 0.961 |
| F285L | NS | 0.942 |
| N308D | NS | 0.771 |
| N308S | NS | 0.728 |
| I309V | NS | 0.257 |
| T411M | NS | 0.322 |
| P491S | NS | 0.808 |
| R501L | NS | 0.937 |
| R501K | NS | 0.889 |
| S502T | NS | 0.746 |
| G503V | NS | 0.925 |
| G503R | NS | 0.925 |
| M504V | NS | 0.910 |
| Q506R | NS | 0.908 |
| Q510R | NS | 0.897 |
| Q510E | NS | 0.890 |
| L560F | NS | 0.237 |
| Y279C | LS | 0.948 |
| Y279S | LS | 0.955 |
| A461T | LS | 0.765 |
| G464A | LS | 0.934 |
| T468M | LS | 0.926 |
| R498L | LS | 0.948 |
| R498W | LS | 0.940 |
| Q506P | LS | 0.960 |
| Q510E | LS | 0.890 |
| Q510P | LS | 0.952 |
| D61H | JMML | 0.900 |
| D61Y | JMML | 0.932 |
| D61V | JMML | 0.893 |
| Y62D | JMML | 0.944 |
| E69V | JMML | 0.817 |
| E69K | JMML | 0.761 |
| A72V | JMML | 0.505 |
| A72T | JMML | 0.483 |
| T73I | JMML | 0.869 |
| E76K | JMML | 0.888 |
| E76G | JMML | 0.876 |
| E76A | JMML | 0.876 |
| E76V | JMML | 0.891 |
| E76Q | JMML | 0.800 |
| R265Q | JMML | 0.622 |
| R289G | JMML | 0.952 |
| S502L | JMML | 0.903 |
| S502T | JMML | 0.746 |
| G503A | JMML | 0.860 |
| G503R | JMML | 0.925 |
| G503V | JMML | 0.925 |
| Q506P | JMML | 0.960 |

*Supplementary Table 1. List of PTPN11 variants along with associated diseases and pathogenicity scores predicted by Mutpred2.*

| Residue 1 (Chain) | Residue 2 (Chain) | (Å) |
|---|---|---|
| 36 (N) | 163 (C) | 7.59 |
| 36 (N) | 161 (C) | 7.55 |
| 37 (N) | 163 (C) | 7.34 |
| 37 (N) | 161 (C) | 7.26 |
| 47 (N) | 449 (P) | 7.24 |
| 46 (N) | 316 (P) | 7.07 |
| 81 (N) | 160 (C) | 7.04 |
| 47 (N) | 448 (P) | 7.04 |
| 47 (N) | 408 (P) | 7.04 |
| 47 (N) | 409 (P) | 7.01 |

*Supplementary Table 2. Top 10 residue pairs exhibiting the largest distance changes in the difference matrix ∆ induced by the D61N variant; N: N-SH2; C: C-SH2; P: PTP.*